\documentclass[dvips]{acta}
\usepackage{supertabular,lscape,epsfig}
\usepackage{amssymb}
\usepackage{amsmath}

\usepackage{graphicx}

\SetPages{0}{0}

\SetVol{0}{2018}

\begin{document}

\begin{Titlepage}
\Title{One new variable candidate and six nonvariable stars at the ZZ~Ceti instability strip}

\Author{Z~s. B~o~g~n~\'a~r$^{1, 2}$,\ \ C~s. K~a~l~u~p$^{1, 3}$ and\ \ \'A. S~\'o~d~o~r$^{1, 2}$}{$^1$Konkoly Observatory, MTA Research Centre for Astronomy and Earth Sciences, Konkoly Thege Mikl\'os \'ut 15-17, H--1121 Budapest\\
$^2$MTA CSFK Lend\"ulet Near-Field Cosmology Research Group\\
$^3$E\"otv\"os University, Department of Astronomy, Pf. 32, 1518, Budapest, Hungary\\
e-mail:bognar.zsofia@csfk.mta.hu, kalupcsilla@gmail.com, sodor.adam@csfk.mta.hu}



\Received{Month Day, Year}
\end{Titlepage}

\Abstract{We present our results on the continuation of our survey searching for new ZZ\,Ceti stars, inspired by the recently launched \textit{TESS} space mission. The seven targets were bright DA-type white dwarfs located close to the empirical ZZ\,Ceti instability strip. We successfully identified one new pulsator candidate, namely PM~J22299+3024, derived detection limits for possible pulsations of four objects for the first time, and determined new detection limits for two targets.}{techniques: photometric -- stars: oscillations -- white dwarfs}

\section{Introduction}
ZZ\,Ceti (or DAV) stars are low-amplitude and short-period white dwarf pulsators with 10\,500--13\,000\,K effective temperatures. Their light-variations are caused by non-radial \textit{g}-mode pulsations in the 100--1500\,s period range. 

Considering the results of ground-based measurements and space-based observations with the \textit{Kepler} telescope, we can distinguish five stages of the cooling of the
ZZ Ceti stars (Hermes et al. 2017). (1) We detect low-amplitude ($\sim$1\,mmag) and short-period (100--300\,s), low-radial-degree pulsations near the blue edge of the instability strip. (2) The periods are still short at a few hundred degrees cooler stage, but the pulsation amplitudes rise up to $\sim$5\,mmag. Some modes are stable enough to investigate evolutionary period changes. (3) Pulsators with the highest amplitudes are located in the middle of the instability strip. Several nonlinear combination frequencies emerge in their Fourier spectrum. Short-term amplitude and frequency variations can be detected. (4) The stars may show irregularly recurring outbursts -- increases in the stellar flux up to 15\% (see e.g. Bell et al. 2017) -- as they cool further.
(5) The longest pulsation periods are detected amongst the coolest ZZ Ceti stars, with low amplitudes at the edge of the instability domain.

In this paper, we present our findings on seven ZZ Ceti star candidates situated either close to the blue or the red edge of the empirical DAV instability strip. We selected our targets from the Montreal White Dwarf Database (MWDD; Dufour et al. 2017)\footnote{http://dev.montrealwhitedwarfdatabase.org/home.html}. It presents the atmospheric parameters from different authors, coordinates, brightnesses in different passbands, and an optical spectrum for the white dwarfs. We chose our targets by effective temperature ($T_{\mathrm{eff}}$), surface gravity ($\mathrm{log}\,g$), and brightness: we looked for new pulsators amongst white dwarfs brighter than 16.5 magnitude. With the presented work, we complemented our search for new white dwarf variable stars for \textit{TESS} (Transiting Exoplanet Survey Satellite; Ricker et al. 2015) observations (Bogn\'ar et al. 2018).   

\section{Observations and data reduction}

We performed the observations with the 1-m Ritchey--Chr\'etien--Coud\'e telescope located at the Piszk\'estet\H o mountain station of Konkoly Observatory, Hungary. We obtained data with an FLI Proline 16803 CCD and an Andor iXon+888 EMCDD camera in white light. The exposure times were 30\,s or 40\,s. We chose these sampling times to provide data with high signal-to-noise ratio. However, note that these are relatively long exposures considering the expected short pulsation periods of the hot ZZ~Ceti stars and candidates like GD\,429, making the detection of the shortest-period pulsations more difficult because of the reduced observed pulsation amplitudes caused by phase smearing. We list the journal of observations in Table~1. 
 
We reduced the raw data frames the standard way utilizing \textsc{iraf}\footnote{\textsc{iraf} is distributed by the National Optical Astronomy Observatories, which are operated by the Association of Universities for Research in Astronomy, Inc., under cooperative agreement with the National Science Foundation.} tasks: they were bias, dark and flat corrected before the performance of aperture photometry of field stars. After photometry, we fitted low-order polynomials to the resulting differential light curves, correcting for low-frequency atmospheric and instrumental effects. This latter smoothing of the light curves did not affect the known frequency domain of pulsating ZZ Ceti stars. Finally, we converted the observation times of every data point to barycentric Julian dates in barycentric dynamical time ($\mathrm{BJD_{TDB}}$) using the applet of Eastman et al.\ (2010)\footnote{http://astroutils.astronomy.ohio-state.edu/time/utc2bjd.html}.

\section{Light curve analysis}

We analysed the measurements with the command-line light curve fitting program \textsc{LCfit} (S\'odor 2012). \textsc{LCfit} has linear (amplitudes and phases) and nonlinear (amplitudes, phases and frequencies) least-squares fitting capabilities, utilizing an implementation of the Levenberg-Marquardt least-squares fitting algorithm. The program can handle unequally spaced and gapped datasets.

We found that six of our targets are not observed to vary (NOV) stars, more specifically, we did not find any significant frequencies in their Fourier transform (FT) suggesting pulsations by the available observations. 

Following our previous work (Bogn\'ar et al. 2018), we calculated the significance levels for the different light curves by computing moving averages of the FTs of the measurements, which provided us an average amplitude level ($\langle {\rm A}\rangle$). We considered a peak significant if it reached or exceeded the 4$\langle {\rm A}\rangle$ level (detection limit). If a target was observed on more than one night, we utilized the FT of all the available data. 

Table~1 also lists the 4$\langle {\rm A}\rangle$ significance levels in mmag units for the NOV objects.
We present the light curves of the NOV stars in Fig.~1, while their Fourier transforms along with the significance levels are plotted in the panels of Fig.~2. We summarize the physical parameters of all targets in Table~2.

For the new ZZ Ceti star candidate PM\,J22299+3024, we were able to derive three significant frequencies by the pre-whitening of its discovery light curve. Table\,3 lists these frequencies with their amplitudes and signal-to-noise ratios (S/N), while Figure~3 shows the light curve and its Fourier transform, respectively. Note that we refer to PM\,J22299+3024 as a candidate variable because so far light variations at this object have been detected by one night of observations only.

\section{Discussion}


Figure~4 shows our targets together with the previously known DAV stars and the empirical boundaries of the  instability strip, derived by Tremblay et al.\ (2015) utilizing 3D spectroscopic analyses of white dwarfs. The boundaries of the ZZ Ceti instability strip are defined according to the equations as follows (Tremblay et al. 2015):
\begin{equation}
(\mathrm{log}\,g)_{blue} = 5.96923 \times 10^{-4} (T_{\mathrm{eff}})_{blue} + 0.52431
\end{equation}
\begin{equation}
(\mathrm{log}\,g)_{red} = 8.06630 \times 10^{-4} (T_{\mathrm{eff}})_{red} -  0.53039
\end{equation}

The typical errors of the published physical parameters are usually around 200\,K and 0.05\,dex in $T_{\mathrm{eff}}$ and $\mathrm{log}\,g$, respectively, which are also plotted in Fig.~4. However, the real external uncertainties are usually higher. The external uncertainties can be estimated from different spectra, distinct atmosphere models and distinct fitting procedures, for example, line profile fitting and whole spectra fitting, which can only be done when the flux calibration and extinction correction are properly performed. The external uncertainties obtained this way can be of the order of
500\,K and 0.1\,dex (S.~O. Kepler, private communication, 2009).

Our results on the different stars are as follows.

\textit{EGGR\,155:} Both Kepler et al.\ (1995) and Gianninas et al.\ (2011) concluded that this star is photometrically constant. Kepler et al.\ (1995) published the corresponding detection level of 2.5\,mma (2.7\,mmag). We were able to further constrain this limit to 2\,mmag.

\textit{GD\,340:} Gianninas et al.\ (2011) found this star to be photometrically constant, but they did not publish any significance level for their result. We derived a 2\,mmag detection limit.

\textit{GD\,429:} Gianninas et al.\ (2006) presented this star as a nonvariable with 0.07\% (0.7\,mma=0.8mmag) detection level. This is the only case where we could not obtain a more strict detection level, as our limit is 2.5\,mmag. Gianninas et al.\ (2011) also mention this object as an NOV star.

\textit{PM\,J18073+0357, TON\,451, WD\,1152+795, PM\,J22299+3024:} We did not find any sign of photometric time series observations on these stars in the literature, thus our measurements on the NOV detection limits and the light variations of PM\,J22299+3024 are new results.  

The detection limit  of TON\,451 is significantly higher than for the other objects. This is both due to the unfavourable weather conditions, especially on the first night of observations (cf. Fig.~1), and the faintness of TON\,451: it was our faintest target with its $\sim$16.5\,mag brightness. Nevertheless, TON\,451 is a promising target situated very close to the red edge of the ZZ\,Ceti instability strip. Although measurements put this star inside the instability domain (see Fig.~3), but considering the problem of the above mentioned external uncertainties in the physical parameter determinations, it may actually lay outside of the strip.

The periods detected in PM\,J22299+3024 (Table~3) are in agreement with its location in the ZZ Ceti instability strip. Variables close to the red edge show lower-frequency light variations than their hotter counterparts, typically longer than 600 seconds.


\MakeTable{lrccrrr}{12.5cm}{Journal of observations. `Exp' is the integration time used, \textit{N} is the number of data points and $\delta T$ is the length of the dataset including gaps. In the comment section, we list the 4$\langle {\rm A}\rangle$ significance levels in parentheses for the NOV targets in mmag unit.}
{\hline
Run & \multicolumn{1}{c}{Date} & Start time & Exp. & \textit{N} & $\delta T$ & Comment\\
 &  & (BJD-2\,450\,000) & (s) &  & (h) & \\
\hline
\multicolumn{6}{l}{EGGR\,155:} & NOV(2)\\
01 & 2017 Dec 13 & 8101.184 & 30 & 425 & 4.55 \\
\\
\multicolumn{6}{l}{GD\,340:} & NOV(2)\\
01 & 2018 May 23 & 8262.351 & 30 & 562 & 5.34 \\
\\
\multicolumn{6}{l}{GD\,429:} & NOV(2.5)\\
01 & 2018 Jan 14 & 8133.296 & 30 & 1055 & 10.12 \\
\\
\multicolumn{6}{l}{PM\,J18073+0357:} & NOV(2.5)\\
01 & 2018 June 11 & 8281.429 & 40 & 233 & 2.80 \\
\\
\multicolumn{6}{l}{TON\,451:} & NOV(6)\\
01 & 2018 Mar 04 & 8182.247 & 30 & 898 & 8.14 \\
02 & 2018 Apr 14 & 8223.277 & 30 & 327 & 3.03 \\
\\
\multicolumn{6}{l}{WD\,1152+795:} & NOV(3)\\
01 & 2018 May 20 & 8259.328 & 30 & 502 & 5.71 \\
\\
\multicolumn{6}{l}{PM~J22299+3024:} & ZZ\,Ceti candidate\\
01 & 2018 July 20 & 8320.346 & 30 & 546 & 5.64 \\
\hline
}

\MakeTable{lcccl}{12.5cm}{Physical parameters of the observed targets. We denoted by \textit{G} at the surface gravity when the source of the original physical parameters are from the database of Gianninas et al.\ (2011). We corrected these $T_{\mathrm{eff}}$ and $\mathrm{log}\,g$ values according to the findings of Tremblay et al.\ (2013) based on radiation-hydrodynamics three-dimensional simulations of convective DA stellar atmospheres. In the other cases the source of the parameters was either Limoges et al.\ (2015) ($L$) or Kawka \& Vennes (2006) ($K$), respectively.}
{\hline
ID & Spectral type & $T_{\mathrm{eff}}$ & $\mathrm{log}\,g$ & \textit{V} mag. \\
 & & (K) & (dex) & \\
\hline
EGGR\,155 & DA & 10\,550 & 8.66$^G$ & 14.35\\
GD\,340 & DA & 10\,350 & 7.98$^G$ & 14.70\\
GD\,429 & DA & 12\,100 & 7.61$^G$ & 14.74\\
PM\,J18073+0357 & DA & 10\,410 & 8.09$^L$ & 15.08\\
TON\,451 & DA & 10\,790 & 7.93$^G$ & 16.48\\
WD\,1152+795 & DA & 10\,810 & 8.23$^K$ & 15.90\\
PM~J22299+3024 & DA & 10\,630 & 7.72$^L$ & 15.89\\
\hline
}

\MakeTable{lrrrr}{12.5cm}{Significant frequencies at the new ZZ Ceti candidate star PM\,J22299+3024.}
{\hline
 & \multicolumn{1}{c}{$f$} & \multicolumn{1}{c}{$P$} & \multicolumn{1}{c}{$A$} & S/N \\
 & \multicolumn{1}{c}{[$\mu$Hz]} & \multicolumn{1}{c}{[s]} & \multicolumn{1}{c}{[mmag]} & \\
\hline
$f_1$ & 935(2) & 1070 & 15.4 & 10.8\\
$f_2$ & 1009(3) & 991 & 7.3 & 5.4\\
$f_3$ & 849(4) & 1178 & 5.8 & 4.7\\
\hline
}


\begin{figure}
\centering
\includegraphics[width=0.70\textwidth, angle=270]{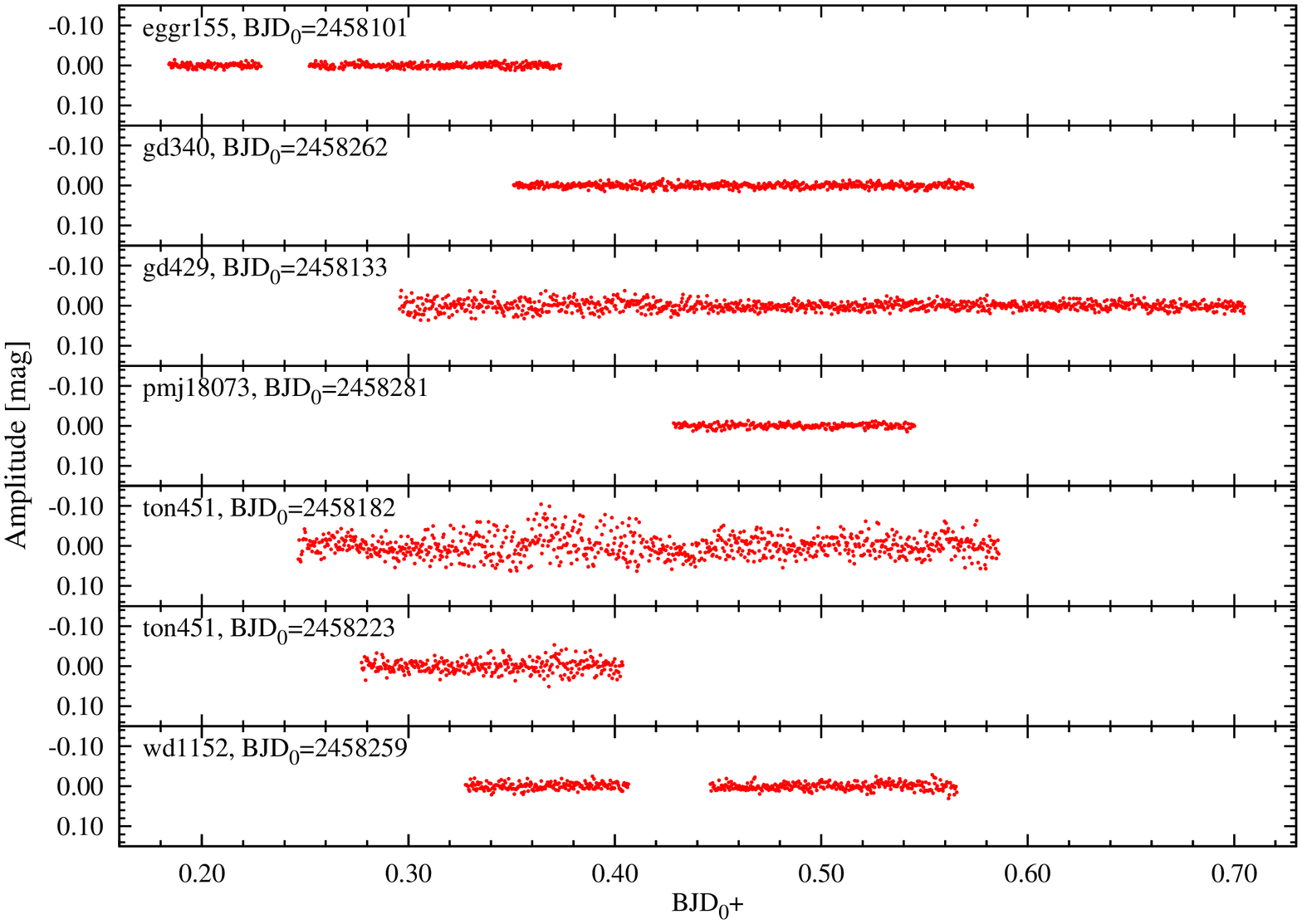}
\FigCap{Normalized differential light curves of the NOV stars.}
\end{figure}

\begin{figure}
\centering
\includegraphics[width=1.0\textwidth]{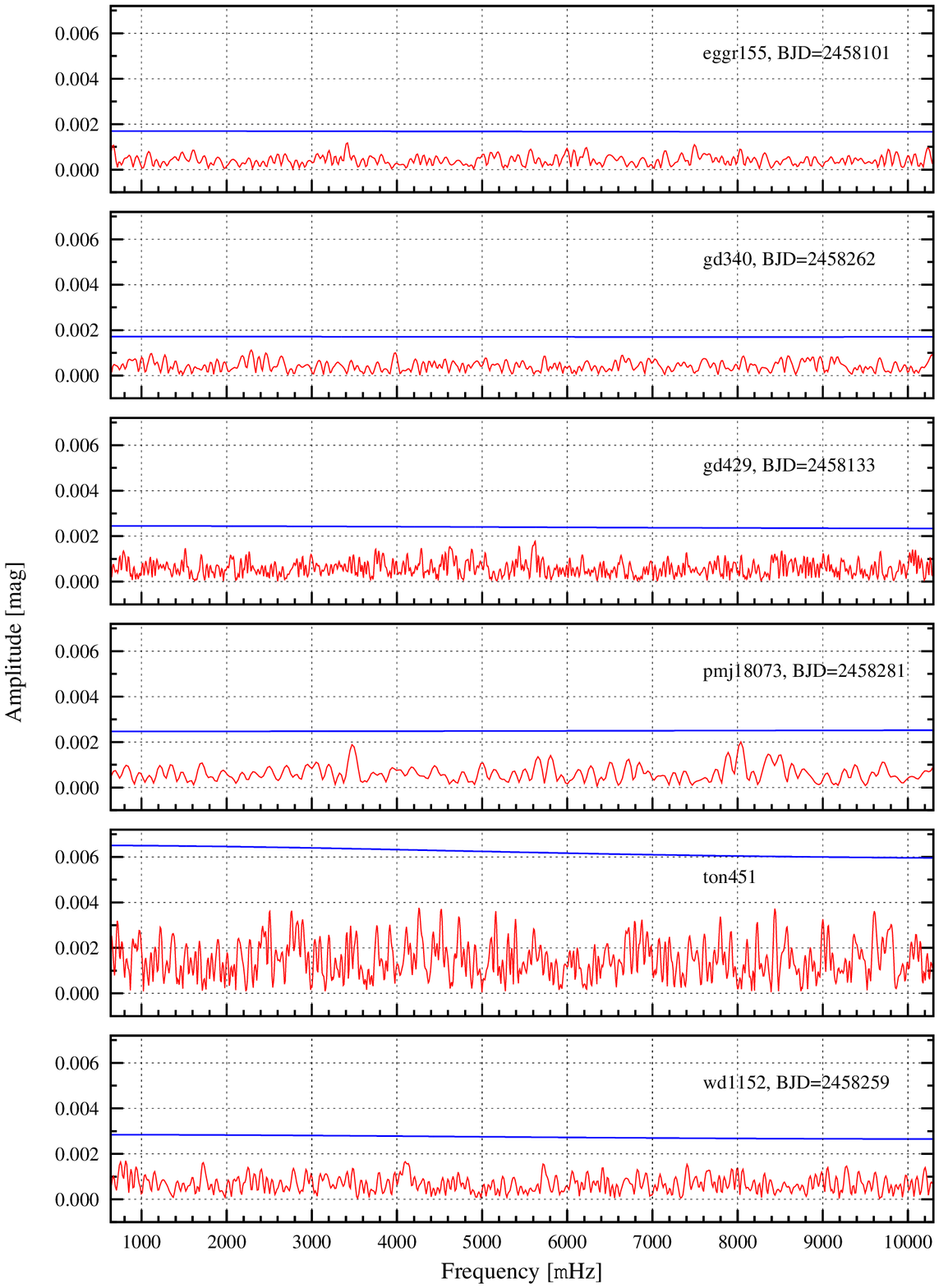}
\FigCap{Fourier transforms of the light curves of the NOV stars. Blue lines denote the 4$\langle {\rm A}\rangle$ significance levels.}
\end{figure}

\begin{figure}
\centering
\includegraphics[width=1.0\textwidth]{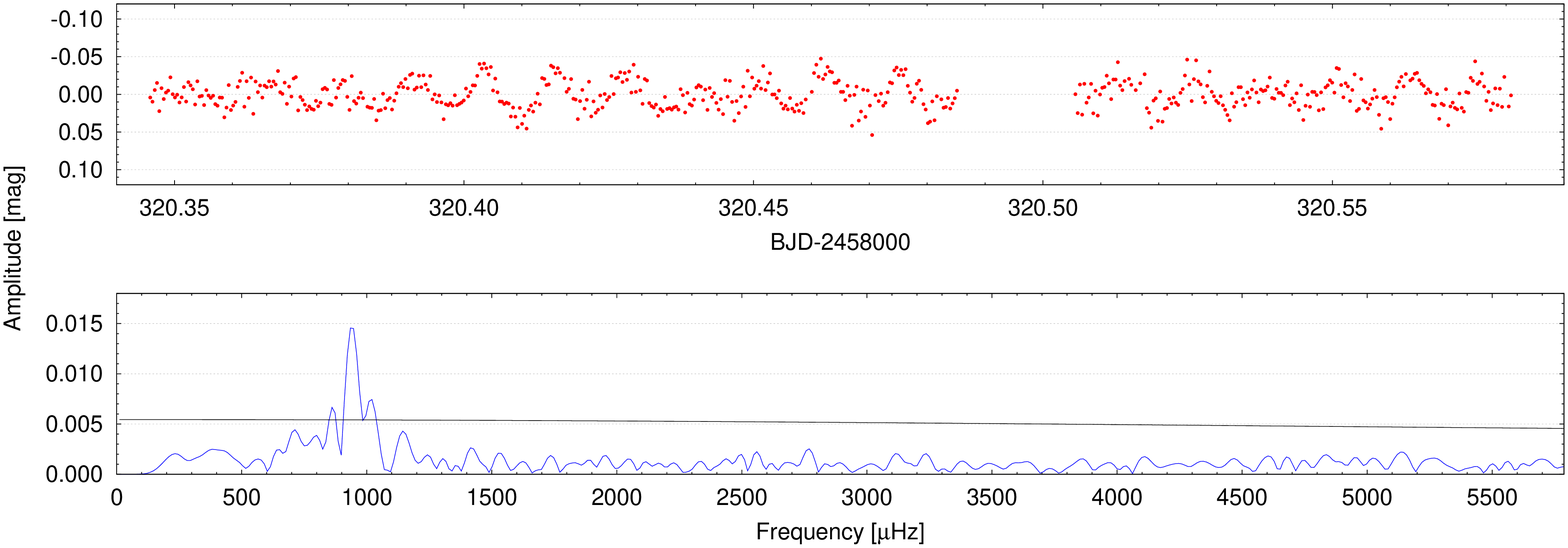}
\FigCap{Light curve and Fourier transform of the of the ZZ Ceti candidate PM\,J22299+3024. Black line denotes the 4$\langle {\rm A}\rangle$ significance level.}
\end{figure}

\begin{figure}
\centering
\includegraphics[width=1.0\textwidth]{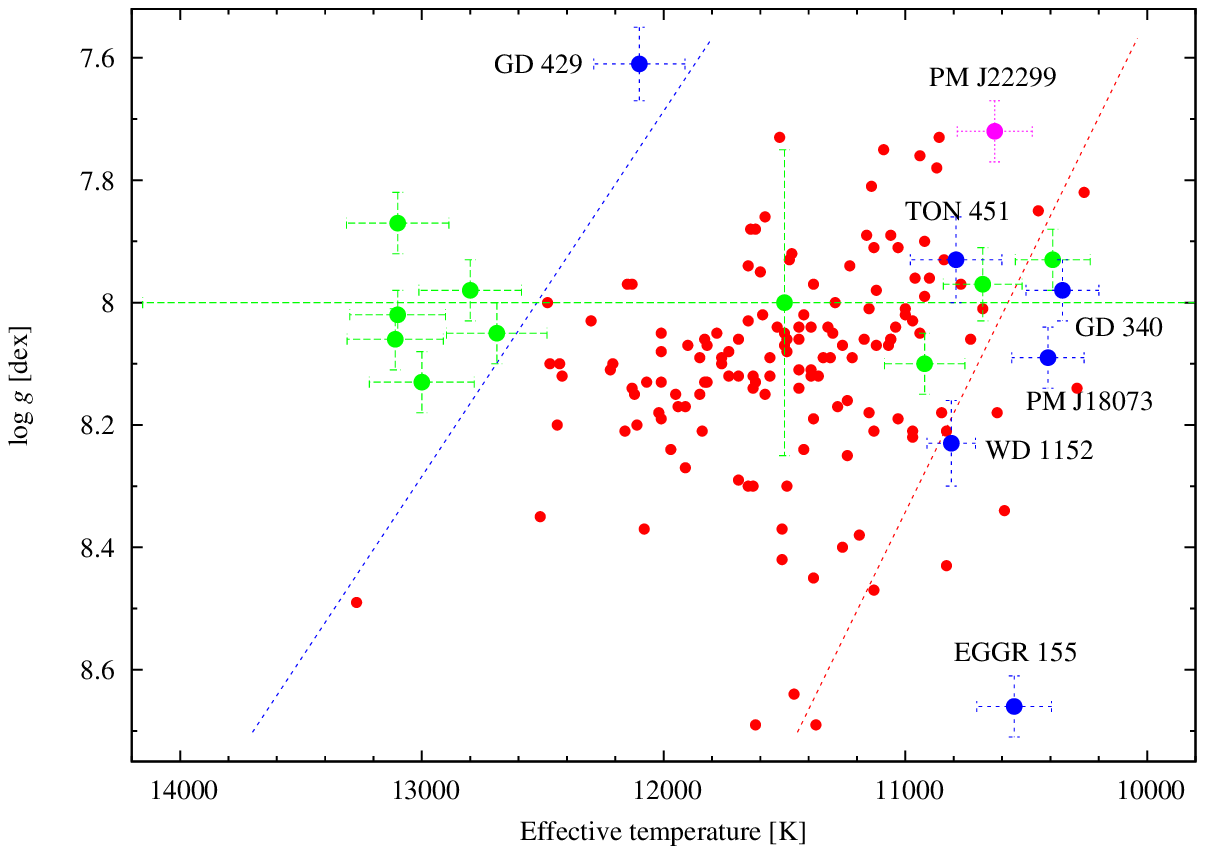}
\FigCap{Known variable stars (red filled dots) and the newly observed ZZ\,Ceti candidates (green, blue and purple dots) in the $T_{\mathrm{eff}}$ -- $\mathrm{log}\,g$ diagram. The atmospheric parameters of the known DAV stars are from the database of Bogn\'ar \& S\'odor (2016). We denoted with green dots the NOV stars listed in Bogn\'ar et al.\ (2018), while blue and purple dots denote the NOV objects and the new ZZ Ceti candidate star presented in this work, respectively. Blue and red dashed lines are plotted at the hot and cool boundaries of the instability strip, respectively, according to Tremblay et al.\ (2015).}
\end{figure}

\Acknow{
The authors thank the anonymous referee for the constructive comments and recommendations on the manuscript. \'AS was supported by the J\'anos Bolyai Research Scholarship of the Hungarian Academy of Sciences, and he also acknowledges the financial support of the Hungarian NKFIH Grant K-113117. \'AS and ZsB acknowledge the financial support of the Hungarian NKFIH Grants K-115709 and K-119517. ZsB acknowledges the support provided from the National Research, Development and Innovation Fund of Hungary, financed under the PD\_17 funding scheme, project no. PD-123910. This project has been supported by the Lend\"ulet grants LP2012-31 and LP2018-7/2018 of the Hungarian Academy of Sciences.
}

\end{document}